\documentclass[reprint,floatfix,amsmath,amssymb,aps,pre,showkeys,showpacs,superscriptaddress]{revtex4-1}
\usepackage[utf8]{inputenc}
\usepackage{graphics}
\usepackage{dsfont}
\usepackage[FIGTOPCAP]{subfigure}
\usepackage{bm}
\usepackage{tikz}
\usepackage[unicode=true, colorlinks,citecolor=blue,breaklinks=true]{hyperref}
\usepackage{natbib}

\usetikzlibrary{shapes,arrows}

\newcommand{\grad}{\vec{\nabla}}
\newcommand{\Dif}[2]{\frac{\partial #1}{\partial #2}}

\newcommand{\ave}[1]{\langle #1 \rangle}
\newcommand{\intr}{\mathcal{I}}
\newcommand{\p}{p}
\newcommand{\vp}{\vec{\p}}
\newcommand{\e}{{\bf \hat{e}}}
\newcommand{\im}{{\rm \bf i}}
\newcommand{\w}{W}
\newcommand{\vw}{\vec{\w}}
\newcommand{\tw}{\tilde{\w}}
\newcommand{\tvw}{\tilde{\vec{\w}}}
\newcommand{\dd}{{\rm d}}

\definecolor{hlcolor}{rgb}{0.0,0.6,0.0}
\definecolor{ntcolor}{rgb}{0,0.6,0}

\newcommand{\req}[1]{Eq.~(\ref{#1})}
\newcommand{\reqs}[2]{Eq.~(\ref{#1}), and~(\ref{#2})}

\begin{document}

\title{Gaussian theory for spatially distributed self-propelled particles}

\date{\today}

\author{Hamid Seyed-Allaei}
\affiliation{Department of Physics, Sharif University of Technology, P. O. Box 11155-9161, Tehran, Iran}
\author{Lutz Schimansky-Geier}
\affiliation{Department of Physics, Humboldt-Universit\"at zu Berlin, Newtonstrasse 15, 12489 Berlin, Germany}
\author{Mohammad Reza Ejtehadi}
\affiliation{Department of Physics, Sharif University of Technology, P. O. Box 11155-9161, Tehran, Iran}
\affiliation{School of Nano Science, Institute for Research in Fundamental Sciences (IPM), P. O. Box 19395-5531, Tehran, Iran}
\email{ejtehadi@sharif.edu}

\begin{abstract}
Obtaining a reduced description with particle and momentum flux densities outgoing from the microscopic equations of motion of the particles requires approximations. The usual method, we refer to as truncation method, is to zero Fourier modes of the orientation distribution starting from a given number. Here we propose another method to derive continuum equations for interacting self-propelled particles. The derivation is based on a Gaussian approximation (GA) of the distribution of the direction of particles. First, by means of simulation of the microscopic model we justify that the distribution of individual directions fits well to a wrapped Gaussian distribution. Second, we numerically integrate the continuum equations derived in the GA in order to compare with results of simulations. We obtain that the global polarization in the GA exhibits a hysteresis in dependence on the noise intensity. It shows qualitatively the same behavior as we find  in particles simulations. Moreover, both global polarizations agree perfectly for low noise intensities. The spatio-temporal structures of the GA are also in agreement with simulations. We conclude that the GA shows qualitative agreement  for a wide range of noise
intensities. In particular, for low noise intensities the agreement with  simulations is better as other approximations, making  the GA to an acceptable candidates of describing spatially distributed self-propelled particles.
\end{abstract}


\maketitle

\section{Introduction}

Study of active matter as an example of non-equilibrium statistical physics has been vastly growing in recent decades~\cite{marchetti,Vicsek2012,Redner2013structure,cates2014,caussin2014,Weber2015random,Solon2015,Ginot2015nonequilibrium,Battle2016broken,Menzel2016on}. There are many examples, such as birds flocks~\cite{Biro2006,Nagy2010,Attanasi2014information,Cavagna2015silent,Cavagna2015flocking}, schools of fishes~\cite{Makris2006,Becco2006,Lopez2012from}, herds of animals~\cite{ginelli2015,toulet2015}, bacteria colonies~\cite{Dombrowski2004self,Sokolov2007concentration,peruani2012}, clusters of cells~\cite{Segerer2015emergence}, and vibrated granular particles~\cite{Narayan2007,kumar2014flocking}. One of the  fundamental and pioneer works in active matter is the introduction of a dynamical microscopic model, known as Vicsek model, to study the emergence of collective behavior~\cite{Vicsek1995}. The result of the model is surprising because it introduces a new class of transitions with continuous symmetry breaking in two dimensions~\cite{chepizhko2013,weitz2015}. Later, many different studies were done to capture other interesting behavior of active matter~\cite{Vicsek1995,Shimoyama1996,Czirok1997,Chate2008a, Peruani2011polar,Farrell2012,solon2013,Weber2014defect,Fily2015dynamics,Bratanov2015new,Chepizhko2015active,nagia2015collective,Grossmann2016superdiffusion,denk2016active}.

Just after the introduction of the Vicsek model, a set of phenomenological continuum equations\footnote{We use the notion continuum equations for a reduced description of the system of self-propelled particles by means of the particle and the momentum flux densities or  the polarization which depend of space and time.},  known as Toner-Tu equations, were proposed~\cite{Toner1995}. The connection between the Vicsek model and the Toner-Tu continuum equations was missing until the equations were derived from a microscopic model of particles with binary collisions using a Boltzmann approach~\cite{Bertin2006,Bertin2009,Peshkov2012,Peshkov2012continuous}. The method was generalized later by considering multiple collisions using an Enskog-type kinetic theory~\cite{ihle2011kinetic,Chou2012kinetic,Ihle2014toward,Ihle2015large}. Similarly, instead of formulating the Boltzmann equation for the probability distribution density equations, one can derive continuum equations using the Fokker-Planck equation~\cite{Baskaran2008,Farrell2012,Grossmann2013}; however, it does not drastically change the result~\cite{Bertin2015}.

One can also write the general form of transport coefficients in terms of the trigonometric moments of the probability distribution, but each moment depends on higher orders~\cite{romanczuk2012swarming,romanczuk2012mean,Grossmann2012}, and still a closure is required. Truncating the series of angular Fourier coefficients of particles distribution is vastly used in active matter to obtain the continuum equations~\cite{Farrell2012,Grossmann2014,Yang2015,allaei2016}. This method has a reasonable accuracy in determining the phase boundaries. Nevertheless, the accuracy of this method for low noise depends on the truncation level. For example, a second order truncation gives transport coefficients that diverge for vanishing noise ~\cite{Farrell2012,allaei2016,Yang2015}. To achieve better results one has to raise the cut-off for the truncation which can be done numerically~\cite{Grossmann2013}. Similar to an expansion of the transport coefficients in terms of trigonometric moments~\cite{romanczuk2012swarming,romanczuk2012mean,Grossmann2012}, we recently tried to obtain transport coefficients in terms of polynomial moments of the distribution, and the closure was done by neglecting higher moments what appeared to be valid for sharp distributions~\cite{allaei2016}. Although one can obtain finite transport coefficients for vanishing noise in this limit, the continuum equations are not able to predict accurately the transition point of the system.

Here we are going to find continuum equations of a simple variant of the Vicsek model within a Gaussian approximation (GA) which is easy to apply and which will predict the system's properties for a wide range of noise intensities. This Gaussian approximation is a systematic approximation since it guarantees the positivity of the probability distributions density. Other approaches to  the hierarchy,  as for example, putting to zero the values of higher moments starting from a certain number might violate the positivity of the density~\cite{Hanggi1980}. From the physical point of view, we assume independence of the locally acting noise and neglect temporal as well as spatial correlations of the fluctuations. Also the fluctuations are generated by a sufficiently large number of independent microscopic degrees of freedom (Gaussianity). Such approximation was used to describe systems of stochastic Kuramoto phase oscillators~\cite{sonnenschein2013approximate,sonnenschein2013excitable,sonnenschein2014cooperative,Sonnenschein2015}, but so far  it  was not applied to the aligning self-propelled particles. In difference to other approximations, we do not truncate in the GA the series of Fourier modes starting with a given number. Alternatively, assuming Gaussian distributed orientations of the velocities at every point, all Fourier modes are considered. The latter are given by the local mean and local variance of the orientations.

The structure of this article is the following. We first introduce the model for which we apply the GA  in Sec.\ref{section:model}. Results of microscopic simulations are presented in Sec.\ref{section:simulations}. Then we formulate a nonlinear Fokker-Planck equation in Sec.\ref{section:kinetic-theory}. The Fokker-Planck equation is used to obtain continuum equations in Sec.\ref{section:hydrodynamics}. Afterward we test the GA in different aspects in Sec.\ref{section:test}, and finally analyze its behavior in Sec.\ref{section:linear-stability}.

\section{Microscopic Model}
\label{section:model}

In our study we use a time continuous Vicsek model that has been introduced before~\cite{peruani2008mean}, and is known as specific case of more general systems~\cite{Farrell2012,Yang2015,Morin2015}. Such a model is in a coarse grained level and without hydrodynamic interactions~\cite{Hernandez-Ortiz2005transport,Underhill2008diffusion}. The model is composed of self-propelled particles moving with constant speed $v_0$ in two dimensions. The orientation of the velocity vector is defined by $\theta$ being the angle between the vector and the $x$-axis.  The direction for motion of the particle is denoted by the unit vector $\hat{v}_\theta$. The rotations of particles depend on the alignment interaction. The dynamics of the particles are given by
\begin{equation}
\dot{\vec{r}}_i = v_0 \hat{v}_{\theta_i} + \sqrt{2K} \vec{\zeta}_i(t),
\label{eq:r-dynamics}
\end{equation}
\begin{equation}
\dot{\theta}_i = \gamma \sum_j F(\theta_j - \theta_i, \vec{r}_j - \vec{r}_i) + \sqrt{2 D_r} \eta_{i}(t),
\label{eq:theta-dynamics}
\end{equation}
where $\vec{r}_i$ is the position of the $i$th particle, and $\hat{v}_{\theta_i}$ indicates unit vector along swimming direction of the particle ($\hat{v}_{\theta} = \cos(\theta_i) \hat{e}_x + \sin(\theta_i) \hat{e}_y$). The noise terms in \reqs{eq:r-dynamics}{eq:theta-dynamics}, $\sqrt{2K} \vec{\zeta}_i(t)$,  and $\sqrt{2 D_r} \eta_{i}(t)$ represent stochastic effects of the environment or/and  of the propulsive mechanism on translational, and rotational movement of particles, respectively. Where $K$, and $D_r$ are the translational and rotational diffusion coefficients. The stochastic functions $\vec{\zeta}_i$, and $\eta_{i}(t)$ are considered to be Gaussian uncorrelated white noises with the properties $\ave{\zeta_i^l} = \ave{\eta_i} = 0$, $\ave{\zeta_i^l(t) \zeta_j^{l^\prime}(t^\prime)} = \delta_{i,j} \delta_{l,l^\prime} \delta(t - t^\prime)$, $\ave{\eta_i(t) \eta_j(t^\prime)} = \delta_{i,j} \delta(t - t^\prime)$ where $l$, and $l^\prime$ indices refer to the vector components of the $\vec{\zeta}_i$. On the right hand side (r.h.s) of \req{eq:theta-dynamics}, $F(\theta,\vec{r})$ is the interaction responsible for an alignment of two particles, and $\gamma > 0$ represents the strength of this alignment. We choose $F(\theta,\vec{r}) = \frac{1}{\pi R^2} \sin \left( \theta \right)$ if $r<R$, and $F(\theta,\vec{r}) = 0$ if $r>R$, where $R$ is the interaction cut-off~\cite{Farrell2012}. The normalization of $F$, to $\pi R^2$ is due to the fact that the average number of particles interacting with a given particle is proportional to the area of the interaction $\pi R^2$.

\section{Simulations}
\label{section:simulations}

\begin{figure}
	\centering
	\includegraphics[width=\columnwidth]{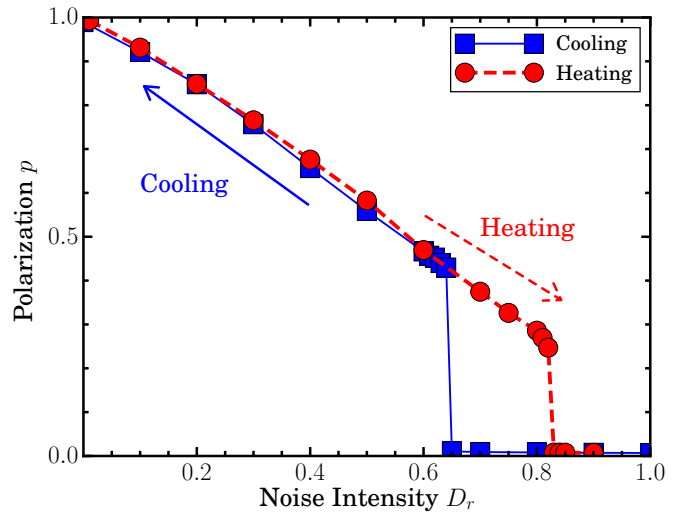}
	\caption{(Color online) Global polarization versus noise intensity for heating and cooling the system. (Red) circles connected by dashed lines are the data points of a heated system, and (blue) squares connected by solid lines corresponds to a cooled system. The (blue) solid arrow and, (red) dashed arrow correspond to the direction of noise change in cooling and heating the system, respectively. Simulation parameters are $\rho_0=8$, $\gamma=\tfrac{1}{8}$, $R=1$, $v_0=1$, $K=\tfrac{1}{8}$, $L_x=128$, and $L_y=32$}
	\label{fig:sim-hystersis}
\end{figure}

One feature of the aligning active particles is the presence of meta-stability and hysteresis~\cite{Gregoire2004,nagy2007,Chate2008,Chate2008a,Ihle2013invasion,solon2015from,solon2015flocking,solon2015pattern,thuroff2014numerical}. We can observe this behavior in a system of particles obeying \reqs{eq:r-dynamics}{eq:theta-dynamics}. In order to show this, we integrated the microscopic \reqs{eq:r-dynamics}{eq:theta-dynamics} inside a periodic box with a time step $dt=\tfrac{1}{64} \tfrac{R}{v_0}$. In simulations, and in all numerical computations, we use units of length and time such that $R=1$, and $v_0=1$. The other parameters where adjusted to $\gamma = \tfrac{1}{8}$, initial density $\rho_0=8$, $K = \tfrac{1}{8}$, and the box dimensions $L_x=128$, and $L_y=32$. Initially the particles are uniformly placed in the space. In the heating process particles are initially aligned in ${\rm x}$ direction, since we are interested in the stability of homogeneous polar state. In the cooling process we orient particles isotropically, because at high noise, the system quickly gets non-polar and homogeneous. At each noise intensity, we wait $2^{14}$ time steps for the relaxation of the polarization, and then $2^{19}$ time steps for sampling. The final configuration in any noise level is used as initial state for the next noise value.

To study the order-disorder phase transition, we define the global polarization vector of the system as,
\begin{equation}
\vp = \frac{1}{N} \sum_{i=1}^N \hat{v}_{\theta_i},
\label{eq:polarization}
\end{equation}
where $N$ is total number of particles. $\p$, the magnitude of the global polarization vector, is used as the order parameter, later on. One can define a similar variable with respect to a subset of particles which are located in a small region around a given position in order to obtain a local polarization. [see \req{eq:local-polarization}].

\begin{figure}
	\centering
	\includegraphics[width=\columnwidth]{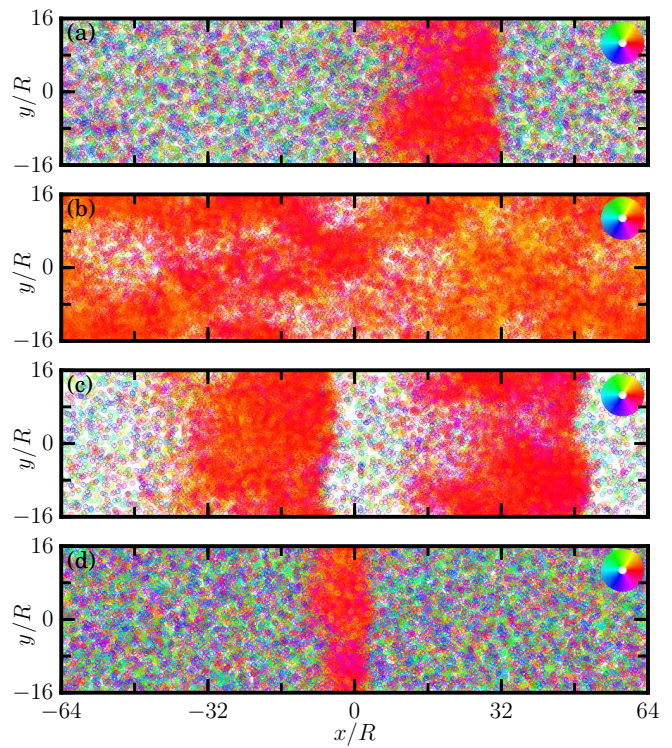}
	\caption{(Color online) Snapshots of the relaxed state of microscopic simulation at different angular noise intensity $D_r$, for (a) cooling, and (b)-(d) heating the system. Particles are shown as small open circles and their color scale indicates their moving direction according to the wheel on top right of each image. (a) Cooling the system from above to $D_r=0.5$, results in a traveling polarized band surrounded by the disordered state. (b) Starting from a homogeneous polar state and heating the system, it remains in the homogeneous polarized state by setting $D_r=0.1$. (c) The system exhibits a sequence of mixed traveling bands between disordered and ordered states from $D_r=0.2$, when we are in heating simulations. (d) At $D_r=0.82$ in the heating process, the number of ordered bands is reduced to a single traveling one in steady state. The band in (d) will vanish if one increases the noise to $D_r=0.83$. Simulation parameters are set to $\rho_0=8$, $\gamma=\tfrac{1}{8}$, $R=1$, $v_0=1$, $K=\tfrac{1}{8}$, $L_x=128$, and $L_y=32$}
\label{fig:snapshots-sim}
\end{figure}

Figure~\ref{fig:sim-hystersis}, shows the hysteresis effect for the cooled and the heated system. Close to the transition points we have chosen smaller changes of the noise intensity. The curve exhibits the same slope which shows the system had in the simulations the sufficient time to relax. In addition to the hysteresis effect, discontinuous transitions are well visible. Both the discontinuous transitions and the hysteresis effect are the results of spatio-temporal band structures found in the system~\cite{Gregoire2004,nagy2007,Chate2008,Chate2008a,Ihle2013invasion,solon2015from,solon2015flocking,solon2015pattern,thuroff2014numerical}. Figure~\ref{fig:snapshots-sim} presents the spatio-temporal structures that are observed in steady state of the system. When we cool the system, it jumps from a homogeneous non-polar state to a mixed state with an ordered band which travels in the background of a gaseous non-polar phase [see Fig.~\ref{fig:snapshots-sim}(a)]. Cooling the system more, will increase the width of the polar band, until it covers the whole space at very low noise level ($D_r < 0.1$). On the other hand, when we start from a homogeneous polar state and heat the system, the homogeneity of the system remains until $D_r<0.2$ [see Fig.~\ref{fig:snapshots-sim}(b)]. At noise levels $0.2 \leq D_r \leq 0.5$, multiple bands are formed [see Fig.~\ref{fig:snapshots-sim}(c) for two bands]. The multiple bands are absorbed into a single band from $D_r=0.6$ [see Fig.~\ref{fig:snapshots-sim}(d)]. By increasing the noise further on, the width of the single band decreases until it disappears and the system gets non-polar.

\section{Kinetic Equation}
\label{section:kinetic-theory}

Let $f(\theta,\vec{r},t)$ be the density of particles at point $\vec{r}$ which move with the angle $\theta$, and $P(\theta,\vec{r},t)$ be the orientational probability distribution of particles at that point. We can simply write the density of particles as
\begin{equation}
\rho(\vec{r},t) = \int_{0}^{2\pi} f(\theta,\vec{r},t) \dd \theta,
\end{equation}
and a relation between $f$,$P$ and $\rho$, that is $f(\theta,\vec{r},t) = \rho(\vec{r},t) P(\theta,\vec{r},t)$. In our system, the most interesting quantities are the local polarization vector,
\begin{equation}
\vp(\vec{r},t) \equiv \ave{\hat{v}_{\theta}} = \int_{0}^{2\pi} \hat{v}_{\theta} P(\theta,\vec{r},t) \dd \theta,
\label{eq:local-polarization}
\end{equation}
and the local momentum flux
\begin{equation}
\vw(\vec{r},t) \equiv \rho \ave{\hat{v}_{\theta}} = \int_{0}^{2\pi} \hat{v}_{\theta} f(\theta,\vec{r},t) \dd \theta.
\label{eq:local-w}
\end{equation}

We aim to derive the continuum equations for the continuous quantities $\rho(\vec{r},t)$, $\vw(\vec{r},t)$ and $\vp(\vec{r},t)$. To this end, with the method presented in reference~\cite{Dean1996}, the Fokker-Planck equation up to second order of spatial derivatives can be derived,
\begin{equation}
\begin{aligned}
\Dif{{f} (\theta,\vec{r},t)}{t} + v_0 \hat{v}_{\theta} \cdot \grad f &= D_r \frac{\partial^2 f}{\partial \theta^2} + K \nabla^2 f -\frac{\partial}{\partial \theta} \intr (\theta,\vec{r},t)
\end{aligned}
\label{eq:simple-fokker-planck}
\end{equation}
Here the second term on the left hand side (l.h.s) shows the advection of particles. On the r.h.s, the first and the second terms come from the rotational and translational diffusion of particles, respectively. $\intr(\theta,\vec{r},t)$ on the r.h.s of \req{eq:simple-fokker-planck} represents the interaction
\begin{equation}
\begin{aligned}
&\intr(\theta,\vec{r},t) = \\
&\gamma \int_0^{2 \pi} \int_{\Omega} F(\theta^\prime - \theta,\vec{r}^\prime - \vec{r}) f^{(2)}(\theta,\vec{r},\theta^{\prime},\vec{r}^\prime,t)  \dd \theta^{\prime} \dd \vec{r}^\prime,
\end{aligned}
\label{eq:exact-int}
\end{equation}
where $f^{(2)}(\theta,\vec{r},\theta^{\prime},\vec{r}^\prime,t)$ denotes the two points density distribution of particles, and the integral over $\vec{r}^\prime$ is done on the subspace $\Omega$, a circle with radius $R$ around the point $\vec{r}$.

To deal with the interacting term of the Fokker-Planck equation [\req{eq:exact-int}] we  neglect correlations of the particles. In this approximation, we factorize $f^{(2)}$ into the product of two single particle density functions. Then we take the integral over $\Omega$ approximately by expanding $f(\theta^\prime,\vec{r}^\prime, t)$ in Taylor series around point $\vec{r}$ up to the second order of the spatial derivatives. The resulting interaction term becomes
\begin{equation}
\begin{aligned}
&\intr(\theta,\vec{r},t) \approx \gamma f(\theta,\vec{r},t) \\
&\Bigg[ \int_0^{2 \pi} \sin(\theta^{\prime} - \theta) \big( f(\theta^{\prime},\vec{r},t) + \frac{R^2 \nabla^2 f(\theta^{\prime}, \vec{r}, t)}{8} \big) \dd \theta^{\prime} \Bigg].
\end{aligned}
\label{eq:fokker-planck-int}
\end{equation}
One can expand $f$ in higher orders of Taylor series, and gets higher orders of spatial derivatives, or neglect the spatial derivatives to obtain a mean-field approximation.

Before constructing the continuum equations, we first find the exact homogeneous mean-field  solution of \req{eq:simple-fokker-planck}. In this aim, we neglect spatial fluctuations by considering homogeneous density of particles $f_H(\theta,t)$. Then \req{eq:simple-fokker-planck} becomes,
\begin{equation}
\begin{aligned}
&\Dif{{f}_H (\theta,t)}{t} = D_r \frac{\partial^2 f_H(\theta,t)}{\partial \theta^2} \\
&-\gamma \frac{\partial}{\partial \theta} \left[f_H(\theta,t) \int_0^{2\pi} \sin(\theta^\prime - \theta) f_H(\theta^\prime,t) \dd \theta^\prime \right].
\end{aligned}
\label{eq:mean-field-fokker-plank}
\end{equation}

Similar to noisy Kuramoto phase oscillators~\cite{Shinomoto1986phase,Sakaguchi1988phase}, and self-propelled particles~\cite{peruani2008mean,Grossmann2013}, the steady state of \req{eq:mean-field-fokker-plank} is given by a von Mises distribution
\begin{equation}
f_H(\theta) = \frac{\rho_0}{2 \pi} \frac{e^{\kappa \p \cos(\theta - \theta_0)}}{I_0(\kappa \p)},
\label{eq:von-mieses}
\end{equation}
where $\p$ is the magnitude of the global polarization vector defined in \req{eq:polarization}, the arbitrary angle $\theta_0$ defines the orientation of the collective motion, $\kappa=\gamma \rho_0 / D_r$, and $I_{\nu}$ indicates modified Bessel function of the first kind. By calculating $\ave{e^{i \theta}}$, one can obtain a self-consistent equation for the determination of $p$,
\begin{equation}
\p = \frac{I_1(\kappa \p)}{I_0(\kappa \p)}.
\label{eq:von-mieses-polarization}
\end{equation}
$\p = 0$, corresponding to an uniform disordered distribution, is always a solution to \req{eq:von-mieses-polarization}. If $\kappa > 2$, then $\p=0$ becomes an unstable solution, and a non-trivial stable solution emerges~\cite{Shinomoto1986phase,Sakaguchi1988phase}. Therefore, $D_c = \gamma \rho_0 / 2$ indicates the critical value for noise intensity, where $\p=0$ if $D_r > D_c$, and $\p \neq 0$ if $D_r < D_c$.

\section{Continuum equations}
\label{section:hydrodynamics}

To build continuum equations from \req{eq:fokker-planck-int}, we write the orientational density in terms of the Fourier components: $f(\theta,\vec{r},t) = \frac{1}{2\pi} \sum_k e^{\im k\theta} \tilde{f}_k(\vec{r},t)$. The $\tilde{f}_k(\vec{r},t)$ for $k=0,\pm 1,$ and $\pm 2$ have relation with the local density 
\begin{equation}
\rho(\vec{r},t) = \tilde{f}_0(\vec{r},t),
\label{eq:f0}
\end{equation}
the local polarization vector as well as the momentum flux [see \req{eq:local-w}],
\begin{equation}
\begin{aligned}
\w_x(\vec{r},t) = \rho(\vec{r},t) \p_x(\vec{r},t) &= \int_0^{2\pi} \dd \theta \cos(\theta) f(\vec{r},\theta,t) \\ &= \frac{1}{2} \left( \tilde{f}_{-1} + \tilde{f}_{1} \right),\\
\w_y(\vec{r},t) = \rho(\vec{r},t) \p_y(\vec{r},t) &= \int_0^{2\pi} \dd \theta \sin(\theta) f(\vec{r},\theta,t) \\&= \frac{1}{2 \im} \left( \tilde{f}_{-1} - \tilde{f}_{1} \right),
\end{aligned}
\label{eq:f1}
\end{equation}
and the local nematic tensor order parameter $\tensor{Q}$,
\begin{equation}
\begin{aligned}
&\rho(\vec{r},t) Q_{x,x}(\vec{r},t) = - \rho(\vec{r},t) Q_{y,y}(\vec{r},t) = \\
&\int_0^{2\pi} \dd \theta \cos(2\theta) f(\vec{r},\theta,t) = \frac{1}{2} \left( \tilde{f}_{-2} + \tilde{f}_{2} \right), \\
&\rho(\vec{r},t) Q_{x,y}(\vec{r},t) = \rho(\vec{r},t) Q_{y,x}(\vec{r},t) = \\
&\int_0^{2\pi} \dd \theta \sin(2\theta) f(\vec{r},\theta,t) = \frac{1}{2\im} \left( \tilde{f}_{-2} - \tilde{f}_{2} \right).
\end{aligned}
\label{eq:f2}
\end{equation}
Thanks to the linear independence of the Fourier basis $e^{\im k \theta}$, we can split \req{eq:simple-fokker-planck} into an infinite set of separate recurrence equations for different $k=0, \pm 1, \pm 2, \dots$ in the Fourier space
\begin{equation}
\begin{aligned}
&\Dif{{\tilde{f}}_k (\vec{r},t)}{t} + v_0 \partial_x \frac{\tilde{f}_{k-1} + \tilde{f}_{k+1}}{2} + v_0 \partial_y \frac{\tilde{f}_{k-1} - \tilde{f}_{k+1}}{2\im} =\\
&-D_r k^2 \tilde{f}_k + K \nabla^2 \tilde{f}_k + \frac{\gamma k}{2} \tilde{f}_{k-1} \left( \tilde{f}_1 + \frac{R^2}{8} \nabla^2 \tilde{f}_1 \right) \\
&-\frac{\gamma k}{2} \tilde{f}_{k+1} \left(\tilde{f}_{-1} + \frac{R^2}{8} \nabla^2 \tilde{f}_{-1} \right). \\
\end{aligned}
\label{eq:fourier-fokker-planck}
\end{equation}
The last two terms on the l.h.s describe advection. At the r.h.s of \req{eq:fourier-fokker-planck}, the first term represents the relaxation of the Fourier modes to the isotropic distribution due to rotational diffusion, and the second term could be thought of the spread of the moment due to the particles diffusion. The third and the fourth terms indicate non-local short range interaction between particles. $\tilde{f}_1$, and $\tilde{f}_{-1}$ define the local polar order parameter, thus interaction is induced by the coupling of polar order to other moments e.g. density. The non-local terms proportional to $R^2 \nabla^2$ are originated from a 1st order local approximation of the non-local forces [see \req{eq:fokker-planck-int}].

To find a closure to \req{eq:fourier-fokker-planck}, one may assume that as $k$ increases, $\tilde{f}_k$ gets smaller. Then one may truncate the equations at a certain $k$ value~\cite{Bertin2006,Bertin2009,Peshkov2012,Peshkov2012continuous}. The truncation method is discussed in subsection~\ref{subsection:truncation}.  Another closure could be obtained based on Gaussianity of the orientation distribution, as we will present it in the following subsection.

\subsection{Gaussian Approximation}
\label{subsection:GA}

We suppose that probability distribution of particles $P(\vec{r},\theta,t)$ is with respect to the orientations a wrapped Gaussian distribution,
\begin{equation}
P^{g}(\vec{r},\theta,t) = \frac{1}{\sqrt{2 \pi} \sigma(\vec{r},t)} \sum_{m=-\infty}^{\infty} e^{-\frac{(\theta-\bar{\theta}(\vec{r},t) - 2 m \pi)^2}{2\sigma(\vec{r},t)^2}},
\label{eq:Wrapped-Gaussian}
\end{equation}
where $\bar{\theta}(\vec{r},t)$, and $\sigma(\vec{r},t)$ are the average, and the variance of $\theta$ at point $\vec{r}$, and time $t$, respectively. The distribution and its derivatives are continuous and periodic in $[-\pi,\pi]$. One should keep in mind that considering an average direction in disordered phase does not contradict with rotational symmetry of the system, since in isotropic phase, $\sigma \to \infty$ and $P^g(\vec{r},\theta,t)$ becomes an uniform distribution. The moment generating function of $P^g(\vec{r},\theta,t)$ is given by
\begin{equation}
\tilde{P}^g_k(\vec{r},t) = \ave{e^{-\im k \theta}} = e^{-\frac{1}{2} \sigma(\vec{r},t)^2 k^2 } e^{-\im \bar{\theta}(\vec{r},t) k}.
\label{eq:moment-generator-Pk}
\end{equation}
The $\tilde{f}^g_k$ has the same form,
\begin{equation}
\tilde{f}^g_k(\vec{r},t) = \rho(\vec{r},t) e^{-\frac{1}{2} \sigma(\vec{r},t)^2 k^2 } e^{-\im \bar{\theta}(\vec{r},t) k}.
\end{equation}
By the interpretation of $\tilde{f}_{0,\pm 1}$ in \reqs{eq:f0}{eq:f1}, $\tilde{f}^g_1(\vec{r},t) = \w_x(\vec{r},t) - \im \w_y(\vec{r},t) = \rho(\vec{r},t) \exp(-\sigma(\vec{r},t)^2/2 - \im \bar{\theta}(\vec{r},t))$. Decomposing $\vw(\vec{r},t)$ into its magnitude $\w(\vec{r},t)$ and direction $\widehat{\w}(\vec{r},t)$, one finds, $\w(\vec{r},t) = \rho(\vec{r},t) \exp(-\sigma^2/2)$, and $\widehat{\w}_x(\vec{r},t) - \im \widehat{\w}_y(\vec{r},t) = \exp(-\im \bar{\theta})$ where $\widehat{\w}_x(\vec{r},t)$, and $\widehat{\w}_y(\vec{r},t)$ are the components of $\widehat{\w}(\vec{r},t)$ along real, and imaginary axes, respectively. Using these identities we write $\tilde{f}^g_k(\vec{r},t)$ in terms of $\w(\vec{r},t)$, $\w_x(\vec{r},t)$, $\w_y(\vec{r},t)$, and $\rho(\vec{r},t)$.
\begin{equation}
\tilde{f}^g_k(\vec{r},t) = \rho(\vec{r},t) \p(\vec{r},t)^{k^2 - k} (\p_x(\vec{r},t) - \im \p_y(\vec{r},t))^k.
\label{eq:Gaussian-fk}
\end{equation}

Solving \req{eq:fourier-fokker-planck} with the help of this Gaussian approximation [see \req{eq:Gaussian-fk}] for $k=0,-1$ and $1$, it gives us the continuity equation,
\begin{equation}
\Dif{{\rho}(\vec{r},t)}{t} + v_0 \grad \cdot \vw(\vec{r},t) - K \nabla^2 \rho = 0,
\label{eq:continuity}
\end{equation}
and the continuum equation of the momentum flux,
\begin{equation}
\begin{aligned}
&\Dif{{\vw}(\vec{r},t)}{t} + \grad. \tensor{\mathcal{J}}_{\vw}(\vec{r},t) = \left[ \frac{\gamma \rho}{2} - D_r - \frac{\gamma W^4}{2 \rho^3} \right] \vw \\
&+ \frac{\gamma R^2}{16} \left[ \rho \nabla^2 \vw - \frac{W^2}{\rho^3} \left( 2 \vw (\vw . \nabla^2 \vw) - W^2 \nabla^2 \vw \right) \right] \\
&+ K \nabla^2 \vw - \frac{v_0}{2} \grad \left[  \rho - \frac{W^4}{\rho^3} \right] ,
\end{aligned}
\label{eq:Gaussian-W}
\end{equation}
where the tensor of the momentum flux current $\tensor{\mathcal{J}}_{\vw}(\vec{r},t)$ is defined as $\tensor{\mathcal{J}}_{\vw} = v_0 \frac{W^2}{\rho^3} \vw \vw$.

The continuity \req{eq:continuity} is composed of advection of the particles and their diffusion. One can think of the diffusion term $K \nabla^2$ as resistance of the system against density fluctuations with compressiblity modulus $K$. On the l.h.s of the \req{eq:Gaussian-W}, there is advection of the momentum flux. It is evident that the advection term in \req{eq:Gaussian-W} is not in the usual form, which means that the system is not invariant under Galilean transformation.  The interpretation of the other terms on the r.h.s of the \req{eq:Gaussian-W} is as follows. The first bracket indicates local relaxation of the momentum flux and it is the necessary non-linearity for the  break of symmetry and originates flocking behavior. All the terms including Laplacian can be considered as diffusion of the momentum flux in space, it is worth to mention that the diffusion of fluctuations of the momentum flux is an-isotropic and it depends on the direction and amount of local momentum flux. The last bracket accounts for the pressure gradient. One can derive a simpler equation from \req{eq:Gaussian-W} close to the disorder-order transition point [see Appendix~\ref{ap:simplification}].

The form of \req{eq:Gaussian-W} is quite similar to the equations derived by Bertin et al.~\cite{Bertin2006,Bertin2009}  and Toner and Tu~\cite{Toner1995} but it contains higher order nonlinear terms. That means instead of $W^2$, and $W^3$ terms, here we have $W^4$, and $W^5$ terms. We will discuss later, that the scaling of polarization with noise in the vicinity of the transition, depends on the order of these non-linear terms. Like the equations derived in~\cite{Bertin2006,Bertin2009} there is no $\grad \grad.\vw$ term in our equation that is allowed by symmetry arguments. Lastly, we can obtain the continuum equations of a binary collision system from \req{eq:Gaussian-W}\cite{Bertin2006,Bertin2009}, by sending $R \to 0$. In this limit, several  terms proportional to $\nabla^2 \vw$ vanish which had their origin in the non-local character of the  interaction.

\subsection{Truncation Method}
\label{subsection:truncation}

At the onset of the transition there exist the scaling behaviors $\tilde{f}_k \sim \epsilon^{|k|}$, $\partial_t \sim \epsilon$, and $\grad \sim \epsilon$~\cite{Bertin2009}. Thus alternatively, one can derive another equation for the momentum density by truncating \req{eq:fourier-fokker-planck}, i.e. $\tilde{f}_k=0$ if $|k| \geq 3$, $\dot{\tilde{f}}_{\pm 2} = 0$, and neglecting all higher order terms than $\epsilon^3$. Truncating \req{eq:fourier-fokker-planck} from $|k| \geq 3$ is sufficient here; however for a nematic aligning particles system, one has to truncate starting from $|k| \geq 5$~\cite{Peshkov2012,Bertin2015}. Applying this closure leads to the continuity \req{eq:continuity} and
\begin{equation}
\begin{aligned}
&\Dif{{\vw}(\vec{r},t)}{t} + \frac{ \gamma v_0 }{16D_r} \left[  5 \vw \grad \cdot \vw + 3 \vw \cdot \grad \vw \right] \\
&= \left[ \frac{\gamma \rho}{2} - D_r - \frac{\gamma^2 W^2}{8D_r} \right] \vw  \\
&+ \left[ K + \frac{v_0^2}{16 D_r} + \frac{\gamma \rho R^2}{16} \right] \nabla^2 \vw - \frac{v_0}{2} \grad \left[ \rho - \frac{5 \gamma \w^2}{16 D_r} \right].
\end{aligned}
\label{eq:truncation-W}
\end{equation}
The second term on the l.h.s can be thought of the advection. On the r.h.s, the first bracket stands for local relaxation of momentum flux, the terms with Laplacian show the spread of momentum flux due to translational diffusion, advection, and alignment. The last bracket can be thought of a pressure gradient. The terms appearing in \req{eq:truncation-W} are quite similar  to the equations derived by Bertin et al.~\cite{Bertin2006,Bertin2009}. We can also see, \req{eq:truncation-W} is similar to \req{eq:Gaussian-W}, but with different nonlinear terms.

\subsection{Homogeneous Solutions to Continuum Theories}
\label{subsec:homogeneous}

The homogeneous steady solutions of \req{eq:Gaussian-W} are either a polar state [$\p = (1 - D_r / D_c)^\frac{1}{4}$] or a non-polar state ($\p = 0$). The stability of these solutions will be discussed in the Sec.~\ref{section:linear-stability}. The polarization of polar state is in the form of $\p = d^\frac{1}{4}$, where $d = 1 - D_r/D_c$. The homogeneous polar solution of the \req{eq:truncation-W} is in the form of $\p=\sqrt{ 2 d (D_r / D_c)}$. This polarization in the vicinity of the transition scales as $\p \sim d^{\frac{1}{2}}$. The exact mean-filed solution [\req{eq:von-mieses-polarization}] has a similar behavior. Also, the same scaling behavior exists in many other active systems~\cite{Toner1995,Bertin2006,Bertin2009,Farrell2012,Bertin2006,Bertin2009,peruani2008mean}. The scaling $\p \sim d^{\frac{1}{2}}$, is a consequence of the local non-linear terms in the form of $\w^3$. However in the GA, a non-linear in the form of $\w^5$ exists and therefore, the homogeneous polarization scales with $d^{\frac{1}{4}}$. The same scaling is observed in the order parameter of phase oscillators in a complex network, when one uses a GA~\cite{sonnenschein2013approximate,sonnenschein2013excitable,sonnenschein2014cooperative,Sonnenschein2015}. If we use the continuum equations in small systems, the scaling is visible. But since in thermodynamic limit - valid for sufficiently large systems - the transition is of first order, the scaling behavior of the homogeneous solution can not be applied.

\begin{figure}
	\centering
	\includegraphics[width=\columnwidth]{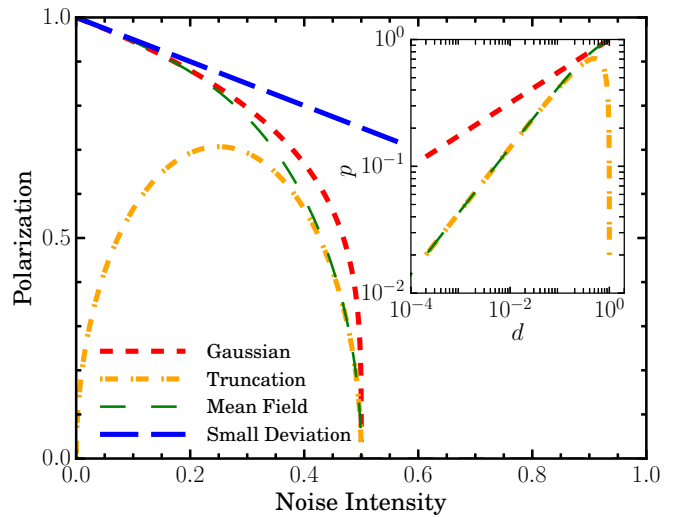}
	\caption{(Color online) Homogeneous polarization of the exact mean-field, the truncation method, the GA, and the small deviation (Eq.~[14] of the Ref.~\cite{allaei2016}), versus noise. All curves, except the small deviation, predict a transition at $D_r=0.5$. The exact mean-field, the GA, and the small deviation converge at low noise $(D_r < 0.1)$. The inset shows the polarization versus $d=(D_c - D_r) / D_c$. One can see that the truncation solution is closer to the exact mean-field solution at the onset of transition, and the GA has polarization greater than both. Finally, within the small deviation approximation, one obtains as homogeneous solution, the blue straight line. It approaches zero at $D_r=2$ which is well beyond $D_c$.}
	\label{fig:mean-field}
\end{figure}

Figure~\ref{fig:mean-field} shows the comparison of the different homogeneous solutions, including the results of the exact mean-field [\req{eq:von-mieses-polarization}], the truncation [\req{eq:truncation-W}], the GA method [\req{eq:Gaussian-W}], and a method with the assumption that the orientation of the particles has very small deviation from the mean value\footnote{Eq.~(14) of the Ref.~\cite{allaei2016}} - The third and higher moments of the distribution are negligible -, which is called small deviation~\cite{allaei2016}. Homogeneous solution of this method is a slowly declining line with a transition point at $D_r=2.0$. The exact mean-field solution and the GA have the same behavior and they coincide in very low noise with the small deviation, but the truncation solution is approaching zero when $D_r \to 0$. In the vicinity of the transition, however, the truncation solution has the same scaling as the exact mean-field solution, but the GA differs [see inset of Fig.~\ref{fig:mean-field}]. the GA is slower in loosing the polarization when it is approaching the transition point. All the homogeneous solutions are predicting a second-order transition, though the transition of the system must be first-order~\cite{Gregoire2004,nagy2007,Chate2008,Chate2008a,solon2015from,solon2015flocking,thuroff2014numerical}. This is due to the spatio-temporal structures in the system that are absent in homogeneous solutions. To obtain the spatio-temporal structures one can either solve the Fokker-Planck \req{eq:simple-fokker-planck}~\cite{thuroff2014numerical,thuroff2013critical}, or the continuum equations~\cite{Peshkov2012,solon2015flocking,solon2015from,allaei2016}. In this article we compare the solutions to different continuum equations which have been obtained from the GA and the truncation method by integrating the partial differential equation in subsection~\ref{subsec:PDE-solution}.

\section{Validity of Gaussian Approximation}
\label{section:test}

\subsection{Distribution of Particles}

\begin{figure}
\centering
\subfigure{\label{subfig:f-alignment-distribution-p-0.1}\includegraphics[width=0.515\columnwidth]{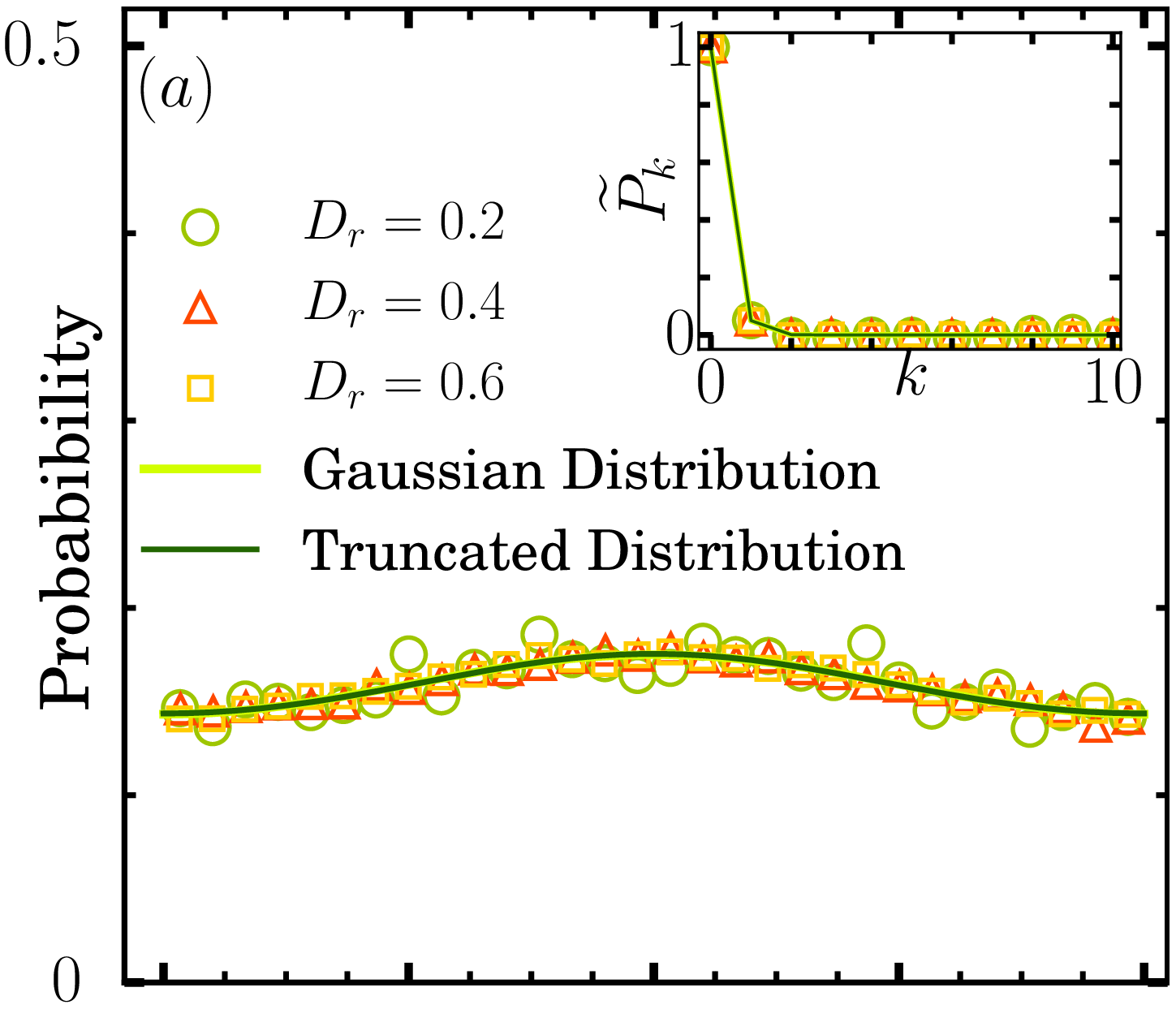}}
\subfigure{\label{subfig:f-alignment-distribution-p-0.4}\includegraphics[width=0.46\columnwidth]{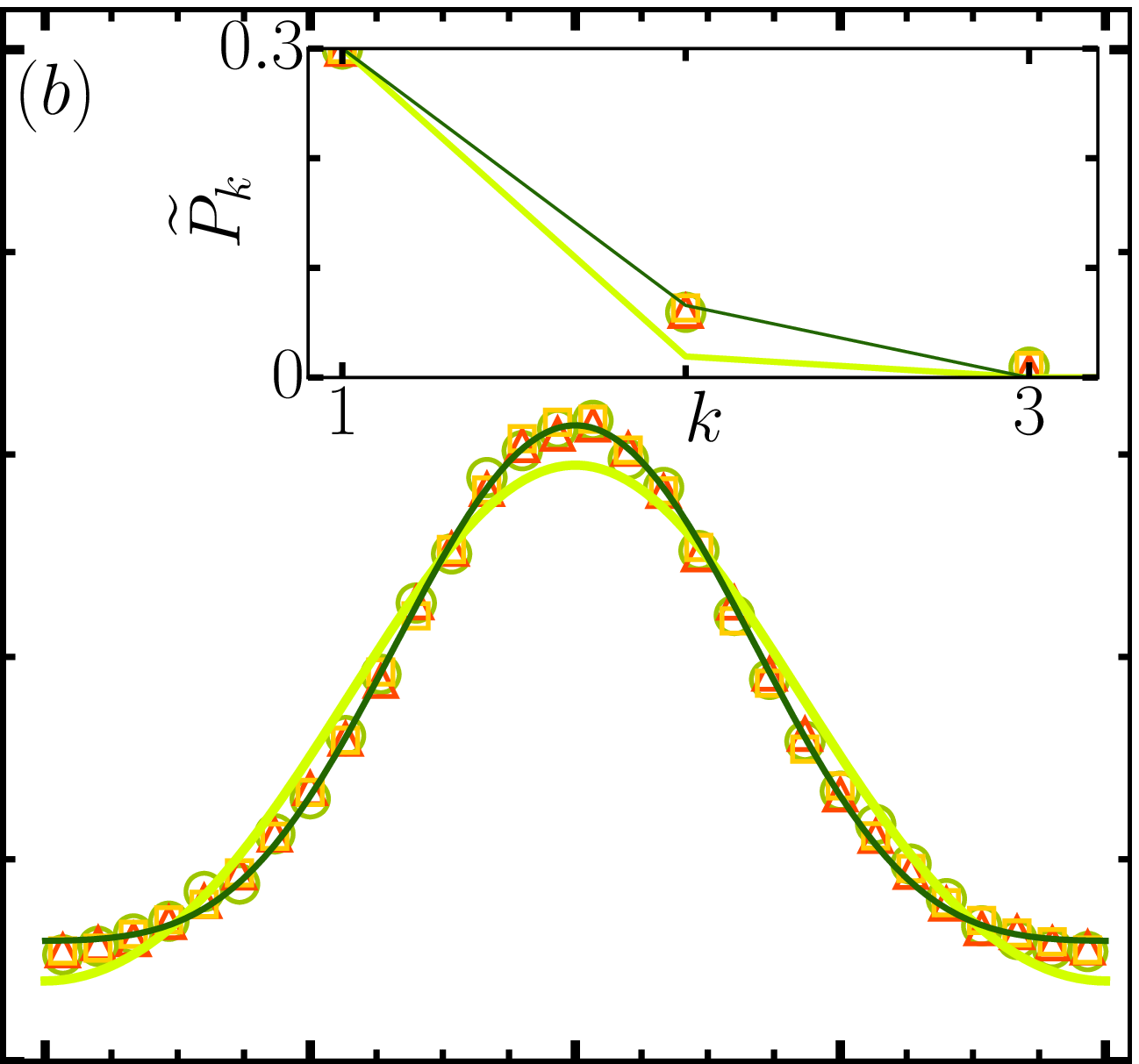}}
\subfigure{\label{subfig:f-alignment-distribution-p-0.7}\includegraphics[width=0.515\columnwidth]{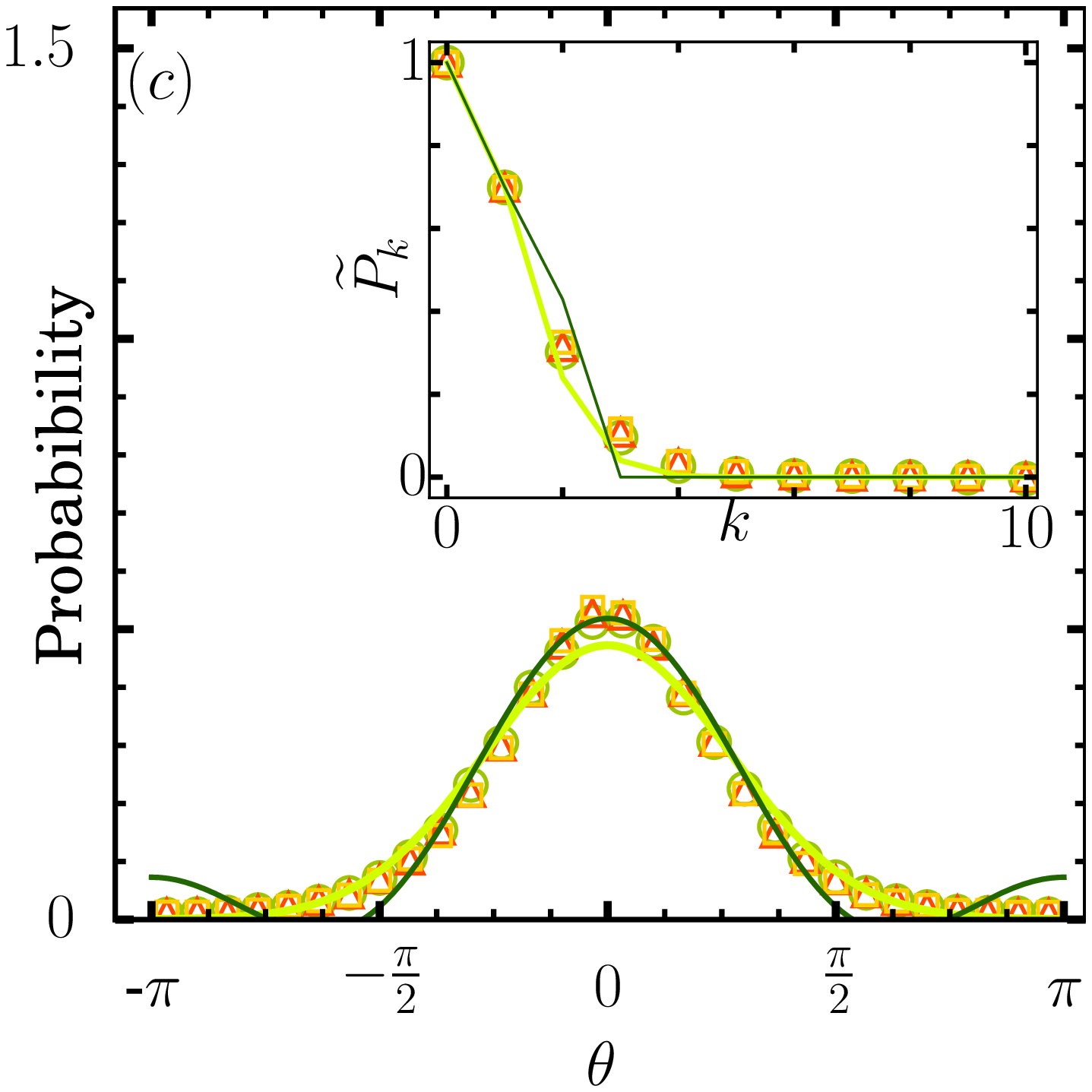}}
\subfigure{\label{subfig:f-alignment-distribution-p-0.9}\includegraphics[width=0.46\columnwidth]{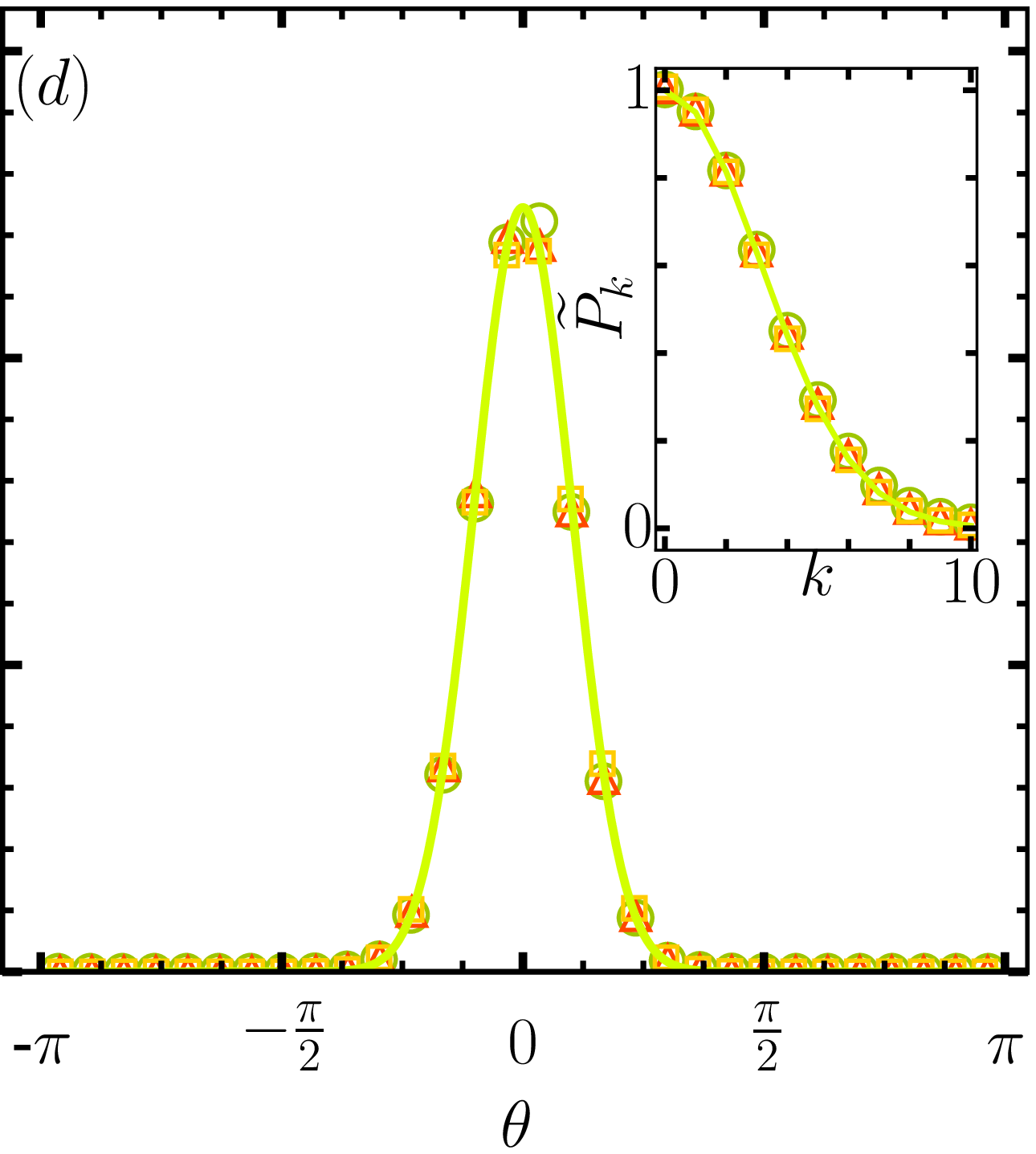}}
\caption{(Color online) Probability distribution function of the orientation of particles for situations with local polarization equal to \subref{subfig:f-alignment-distribution-p-0.1} $0.05$, \subref{subfig:f-alignment-distribution-p-0.4} $0.4$, \subref{subfig:f-alignment-distribution-p-0.7} $0.7$, and \subref{subfig:f-alignment-distribution-p-0.9} $0.95$. The points are obtained from simulation data with noise intensities $D_r=0.2,0.4$, and $0.6$. To obtain the data points we split the box into cells and extracted the mean-centered particles angles of any cell with the corresponding local polarization in an interval of $0.01$. The light green curves are the expected wrapped Gaussian distribution obtained from the theory, and the dark green curves are the truncated distribution at $k=3$ ($\tilde{P}_k = 0$ if $k \ge 3$). The truncated distribution for $p_s > \sqrt{0.5} \approx 0.7$ is not available. Insets represent the generating function $\tilde{P}_k$ of the distributions. Simulation parameters are $\rho_0=8$, $\gamma = \frac{1}{8}$, $R=1$, $v_0=1$, $K=\frac{1}{8}$, $L_x=128$, and $L_y=32$.}
	\label{fig:f-alignment-distribution}
\end{figure}

To confirm the GA we need to show that the distribution of the individual orientation of the particles is close to a wrapped Gaussian distribution. Figure~\ref{fig:f-alignment-distribution} shows the angle distribution $P(\theta|\p_s)$ from particle simulations, that shows the orientation distribution of particles inside cells with size $2R\times 2R$ which has local polarization, $\p_w$, with the condition $|\p_w - \p_s| < 0.01$. $\p_w$ is computed similar to \req{eq:polarization} for the particles inside the cell. In Fig~\ref{fig:f-alignment-distribution}, $P(\theta|\p_s)$ is compared with the theoretical wrapped Gaussian distribution $P^g(\theta|\p_s)$ with the variance given by $\p_s$, $\sigma= \sqrt{-2 \ln(\p_s)}$, and also with the distribution obtained from the truncation method. Each plot corresponds to an ensemble of the particles with $\p_s=0.05,0.4,0.7,0.95$. Next, the distribution of all particles in the ensemble is plotted. This procedure at each $\p_s$ is done for systems with different noise levels ($D_r=0.2,0.3$, and $0.6$). It is evident from the plots that the result of the distribution does not depend on the noise intensity but only depends on $\p_s$. To compare with our theory we plotted the wrapped Gaussian function corresponding to the selected polarization $\p_s$, also we plotted the truncated distribution for the same value of the polarization. One can see that without any free parameter we achieved an astonishing agreement in Fig.~\ref{fig:f-alignment-distribution}. Nevertheless, the truncation distribution fits better for $\p_s=0.4$ [see Fig.~\ref{fig:f-alignment-distribution}(b)].

Figure~\ref{fig:f-alignment-distribution} compares $P(\theta|p_s)$ at different noise levels with the theoretical wrapped Gaussian distribution $P^g(\theta|p_s)$ with the variance given by $\sigma= \sqrt{-2 \ln(\p_s)}$, and also with the distribution obtained from the truncation method. Each plot corresponds to an ensemble of the particles with local polarization equal to $\p_s=0.05,0.4,0.7,0.95$. It is evident from the plots that the result of the distribution does not depend on the noise intensity but only depends on $\p_s$. To compare with our theory we plotted the wrapped Gaussian function corresponding to the selected polarization $\p_s$, also we plotted the truncated distribution for the same value of polarization. 

We also compared the moment generating function of simulation distribution $\tilde{P}_k$, with the corresponding Gaussian moment generating function $\tilde{P}^{g}_k$ in the inset of each plot in Fig.~\ref{fig:f-alignment-distribution}. By a closer look, we can see that for $\p_s = 0.4$, the GA predicts $\tilde{P}^g_2= 0.0256$, that is smaller than the simulation data point $\tilde{P}_2 = 0.0845$, while a truncation at $k=3$ predicts a closer value of $0.0877$ [see inset of Fig.~\ref{fig:f-alignment-distribution}(b)]. Thus we can conclude that in low ordered regions, the truncation approach is the better approximation. In the opposite, the GA for situations with high order ($\p_s=0.95$) considers the asymptotic behavior of the tail of $\tilde{P}_k$. It is also worth to note that the truncation distribution is available only for $\p_s < \sqrt{\frac{1}{2}} \approx 0.7$. Therefore, the GA gives a better agreement in case of low noise.

\subsection{Numerical Solution of the Continuum Equations}
\label{subsec:PDE-solution}

We found inhomogeneous numerical solutions for the continuum equations of both approaches, the GA and the truncation approach [Eq.~\eqref{eq:continuity} to \eqref{eq:truncation-W}]. To determine the binodal and spinodal curves of the first order transition~\cite{solon2015from}, we added a noise to the r.h.s of the \reqs{eq:Gaussian-W}{eq:truncation-W} in the form of $D_r \rho (1-\w^2) \vec{\xi}(\vec{r},t)$, where $\vec{\xi}$ is a non-correlated stochastic field with zero mean and $\ave{\xi_i(t,\vec{r}) \xi_j(t^\prime,\vec{r}^\prime)} = \delta_{ij} \delta(\vec{r} - \vec{r}^\prime) \delta (t - t^\prime)$. The noise is proportional to $D_r$, since it is originated from the stochastic sources of the microscopic equations. 

\subsubsection{Integration Method}

We used pseudo spectral method, semi-implicit time stepping and anti-aliasing techniques $2/3$ rule~\cite{uecker2009short,canuto1993spectral} to integrate the continuum equations with periodic boundary condition. The implicit stepping parameter, $\eta$, is set to $\eta=0.5$, and time steps are equal to $dt=R/32v_0$. We set the unit length, and time such that $R=1$, and $v_0=1$. The system parameters were adjusted as simulations to $\rho_0=8$, $\gamma=\tfrac{1}{8}$, $R=1$, $v_0=1$, $K=\tfrac{1}{8}$, $L_x=128$, and $L_y=32$. The grid dimensions of the integration were chosen $N_x=128$, and $N_y=32$.

In the heating process, we started by a homogeneous polar state with the initial condition $\vec{p}(\vec{r},0) = {\bf \hat{e}}_x$. In the cooling process, we started by a homogeneous non-polar state with initial condition $\vec{p}(\vec{r},0) = \vec{0}$. We then integrate both systems at each noise level for a duration between $2^{12}$ to $2^{15}$ units of time depending on the relaxation time of the polarization at that level. For each noise level, the last $1/16$ of the integration is devoted for computing the polarization average value.

We were faced with a numerical instability during the cooling at $D_r=0.2$. At the onset of the divergence of the continuum fields, a broad polar band had occupied half of the system. The rest of the system had a very low density. Such a configuration with large gradients of the continuum variables is vulnerable for the accuracy of the numerical integration. The occurring inaccuracy during the integration generates negative densities in the neighborhood of the band which causes the mentioned divergence of the fields. Like in the microscopic simulations we expect that the polar band width increases by lowering the noise and the band will occupy homogeneously the whole system. But this situation was not available during the numerical integration of the continuum equations.

To overcome this problem we homogenized the system. To do this, at $D_r=0.2$, we increased the translational diffusion to $K=5$, and integrated for $128$ units of time. In fact, this increase originates a new spatial initial condition. Starting then with this homogeneous field, the instability
disappeared. After integrating $128$ units of time, we put $K$ back to $1/8$, and integrated for $2^{16}$ units of time in order that the system can relax with the correct $K$-value. During the last $2^{11}$ units of time of the relaxation we sampled the system and  computed the wanted value of the global polarization. To justify the procedure, we have proven by long simulations that during the homogenization and the later relaxation, the value of the global polarization $\p$ has not changed. Definitely, the temporall increase of $K$ changes the configuration of the system from a band to a homogeneous state, but the trend of  the computed values of the global polarization in Fig.\ref{fig:f-alignment}(a) and its noise dependence are not affected.

\subsubsection{Integration Results}

\begin{figure}
	\centering
	\subfigure{\label{subfig:p-noise-cooling}\includegraphics[width=\columnwidth]{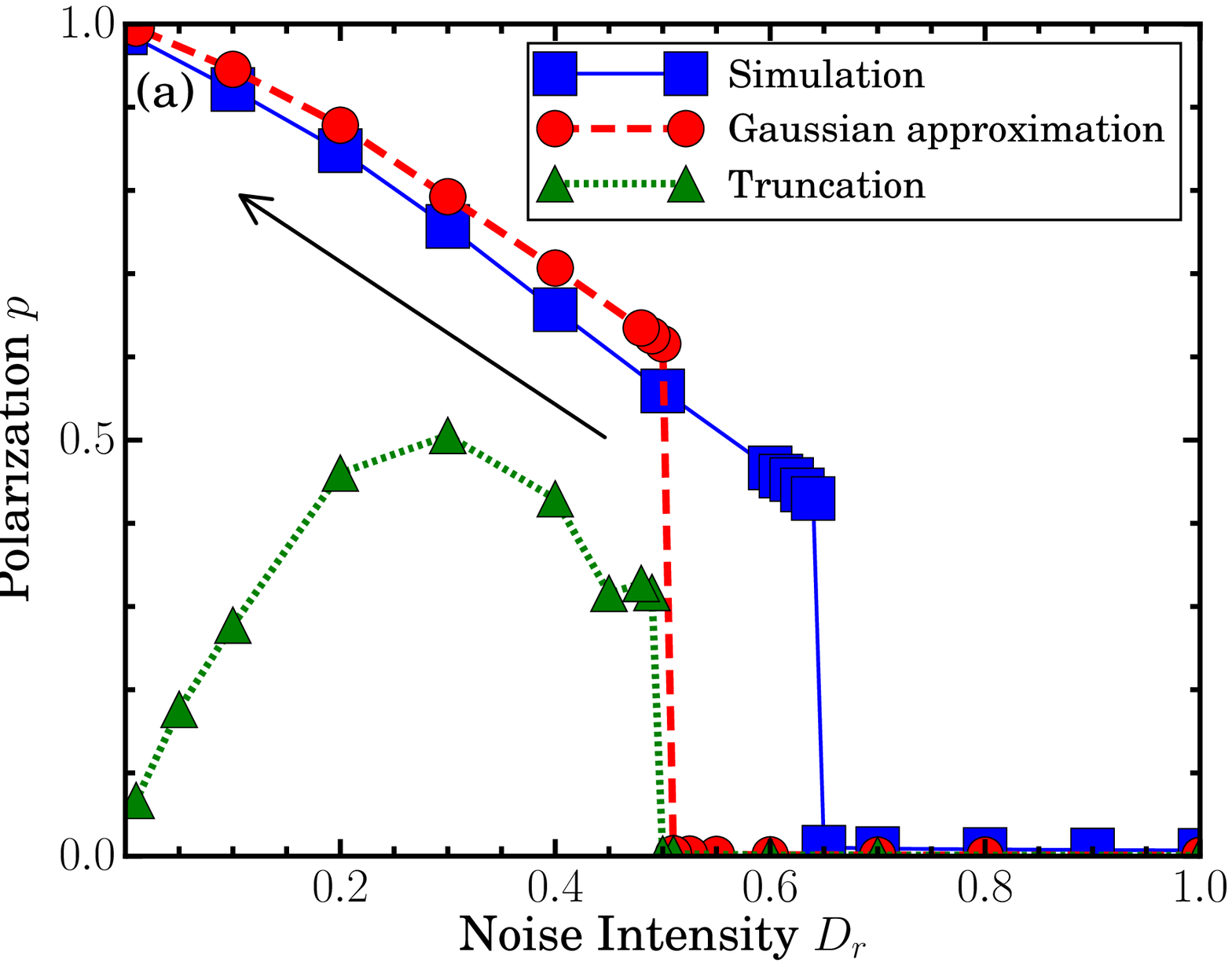}}\\
	\subfigure{\label{subfig:p-noise-heating}\includegraphics[width=\columnwidth]{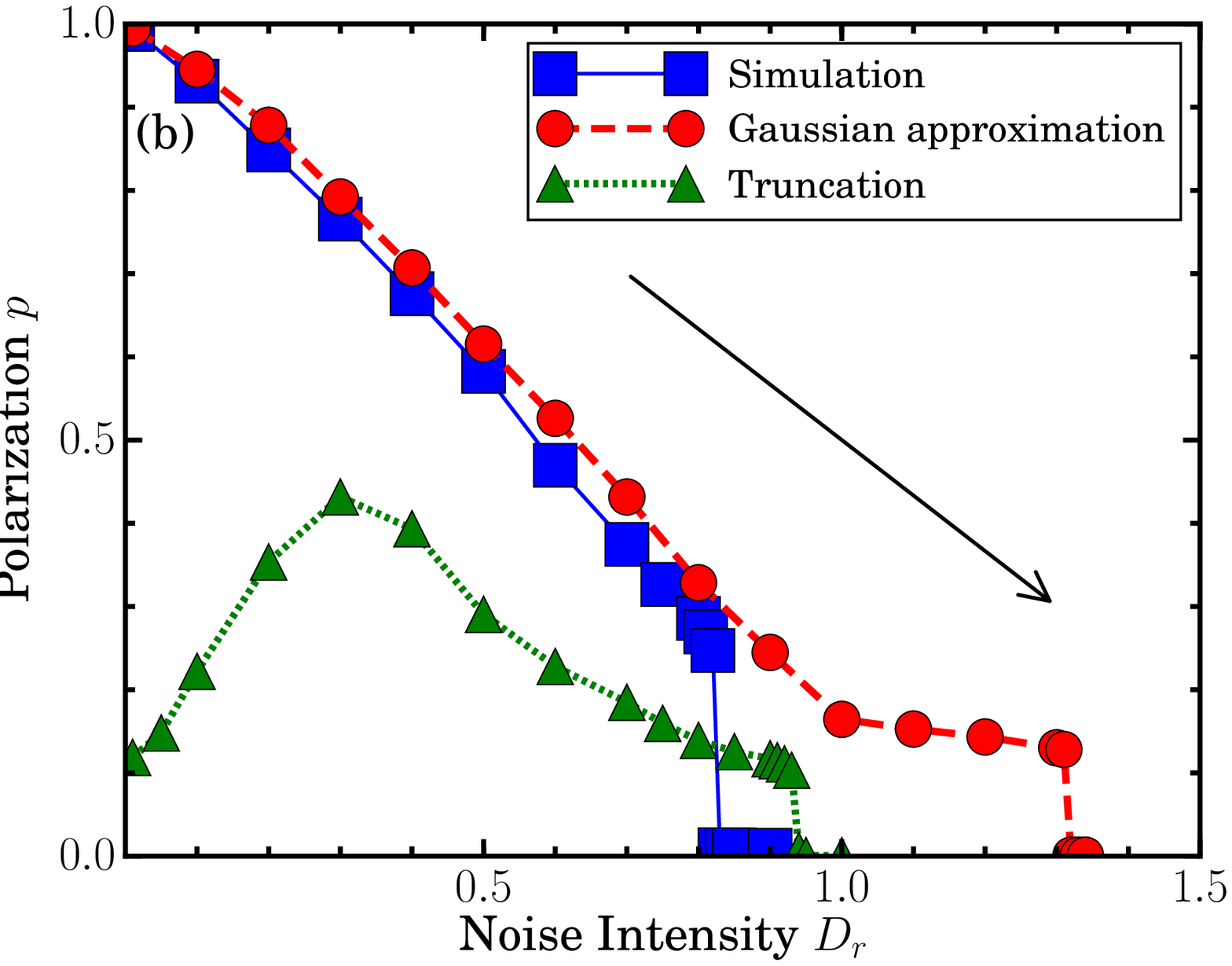}}
	\caption{(Color online) Global polarization, versus noise intensity, $D_r$, of continuum equations, in \subref{subfig:p-noise-cooling} Cooling the system, and \subref{subfig:p-noise-heating} heating the system. The (blue) squares correspond to the result of simulation, the (red) circles correspond to the GA, and the (green) triangles correspond to the truncation method. The arrow in each plot represents the direction for change of the noise. The grid dimension for numerical integration is 128 by 32. Parameters are $\rho_0=8$, $\gamma=\tfrac{1}{8}$, $R=1$, $v_0=1$, $K=\tfrac{1}{8}$, $L_x=128$, and $L_y=32$}
	\label{fig:f-alignment}
\end{figure}

Figure~\ref{fig:f-alignment} compares the global polarization, defined in \req{eq:polarization}, of the GA, the truncation technique, and the simulations versus noise intensity in the cooling and heating procedures. In contrast to the homogeneous solutions, the transitions are discontinuous in both continuum theories and the hysteresis effect is visible. One can see that the GA gives a better global polarization in comparison to the truncation. Nevertheless, the transition point of the GA during heating is beyond the value obtained by microscopic simulation, and the truncation method predicts a closer transition point [see Fig.~\ref{fig:sim-hystersis}].

\begin{figure}
	\centering
	\includegraphics[width=\columnwidth]{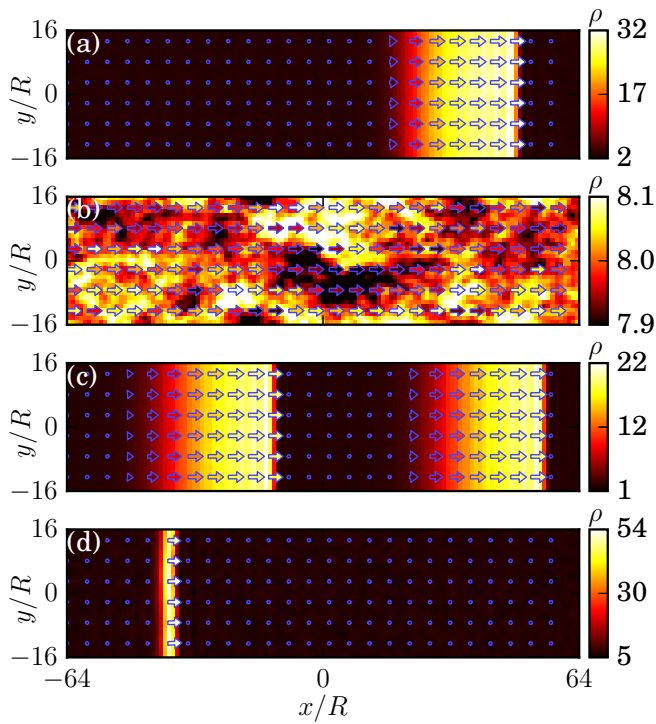}
	\caption{(Color online) Snapshots of the final states of the integration of the GA continuum equations for (a) cooling, and (b)-(d) heating processes. The different colors show the density according to the color bar. Arrows indicate the local polarization vector. The grid dimensions are $N_x=128$, and $N_y=32$. (a) A system after cooling down to $D_r=0.5$ exhibits a traveling band. (b) A heated system at $D_r=0.2$ stays in homogeneous polar state. (c) Heated up to $D_r=0.3$, the system forms wave trains. (d) After heating the system up to $D_r=1.31$ a single narrow traveling band remains in the system. This band will vanish if one increases the noise to $D_r=1.32$. System parameters are set to $\rho_0=8$, $\gamma=\tfrac{1}{8}$, $R=1$, $v_0=1$, $K=\tfrac{1}{8}$, $L_x=128$, $L_y=32$}
\label{fig:snapshots-ga}
\end{figure}

In addition to the hysteresis behavior, one has to check the spatio-temporal structures of the GA continuum equations (see supplementary materials for movies \footnote{See Supplemental Material at [URL will be inserted by publisher] for movies of numeric integration of the GA continuum equations.}). Figure~\ref{fig:snapshots-ga} shows snapshots of the spatio-temporal structures of the GA. These snapshots are obtained after sufficiently waiting for the relaxation of the polarization at an specific value of $D_r$. In the integration of the GA, when we cool the system from above, it jumps from a homogeneous non-polar state, and stays in a mixed state with an ordered band which travels in the background of a disordered phase at $D_r=0.5$ (see Fig.~\ref{fig:snapshots-ga}[a] and \href{SI1-cooling-GA.mp4}{the first movie} in the supplementary materials~\cite{Note1}). On the other hand, when we start from a homogeneous polar state and heat the system, the homogeneity of the system remains until $D_r=0.26$ [see Fig.~\ref{fig:snapshots-ga}(b)]. Slightly above ($D_r=0.27$), the homogeneity of the system breaks down by formation of a wave train. The wave train is not the steady state of the system and the bands in the wave train merge together with a very slow coarsening dynamics~\cite{solon2015pattern}. After a long integration, we observe two bands in the system at $D_r=0.3$ [Fig.~\ref{fig:snapshots-ga}(c)]. In higher noise intensities, the bands are narrower. Moreover, only a single band remains in the system above $D_r=0.75$ [see Fig.~\ref{fig:snapshots-ga}(d)]. The single band disappear at $D_r=1.32$, and the system is in the non-polar disordered state. As one can see, the spatio-temporal structures of the microscopic simulation in Fig.~\ref{fig:snapshots-sim} is quite similar to the results of the GA in Fig.~\ref{fig:snapshots-ga}. The only difference is the noise intensity at which the system configuration change e.g. from homogeneous to an in-homogeneous configuration. This could be related to higher polarization predicted by the GA [see Fig.\ref{fig:f-alignment}].

It has been shown with a dynamical system approach, that the wave train, narrow and wide single band states are the three solutions of the continuum equations as, for example, used in~\cite{caussin2014}. We observe the same solutions depending on the cooling or heating processes. Similar structures are obtained by directly integrating the Boltzmann equation of a binary collision system as well~\cite{thuroff2014numerical}. Th{\"u}roff et al. also found a new pattern formation that is consist of parallel lanes of polar clusters moving against one another. The numerical solution of the continuum equations of the GA does not give cluster-lanes. This could be due to the continuum assumptions that we made, or the difference between the binary and the continuous-time microscopic interaction.

In addition to the solutions that are discussed, one can find more featured behaviors by adding repulsion to the microscopic model of the particles, e.g. smectic phase~\cite{Menzel2013unidirectional,Romanczuk2016emergent,Menzel2016on,Chen2013universality,Adhyapak2013live}. Building continuum equations from microscopic model that describe the smectic phase is chalanging, because the correlations between the particles positions are important, and therefore, applying the GA to such models, requires detailed future investigations and considerations as we have done in the present study.

\section{Linear Stability}
\label{section:linear-stability}

\begin{figure}
	\centering
	\includegraphics[width=\columnwidth]{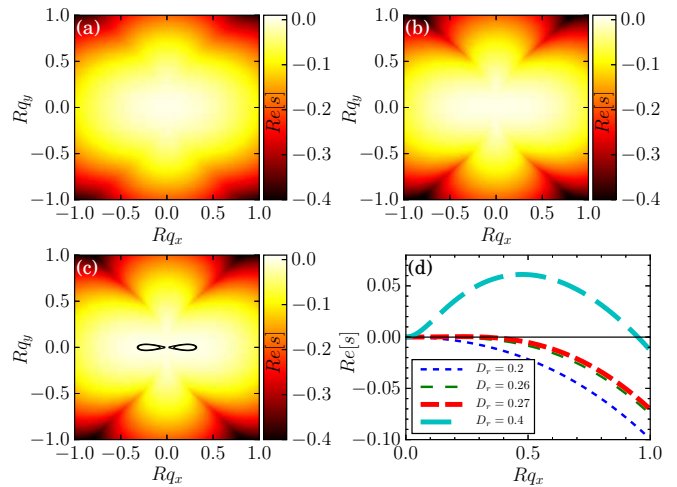}
\caption{(Color online) (a)-(c) Maximum real part of eigenvalues of the perturbations. (a), (b), and (c) corresponds to $D_r=0.2$,$0.26$, and $0.27$ respectively. The color shows the $Re[s]$. The black curves in (c) represent the contours for roots of $Re[s]$ according to the color bars. Inside the contours $Re[s] > 0$. (d) $Re[s]$ for longitudinal waves as a function of wave number $q$ for different values of $D_r$. We see from the figure that close to $D_r=0.27$, the homogeneous polar state is unstable and the system forms in-homogeneous structures. The parameters are set to $\rho_0=8$, $\gamma=\tfrac{1}{8}$, $R=1$, $v_0=1$, $K=\tfrac{1}{8}$.}
\label{fig:s-map}
\end{figure}

In order to find the boundary of behavioral changes of the continuum system, we can study the stability of perturbations in \reqs{eq:continuity}{eq:Gaussian-W} around homogeneous polar state, or disordered phase. For this purpose, we write the fields $\vw$ and $\rho$ as
\begin{equation}
\vw(\vec{r},t) = \w_0 \e_x + \delta \vw(\vec{r},t), \rho(\vec{r},t) = \rho_0 + \delta \rho(\vec{r},t)
\end{equation}
where $\rho_0$, and $\w_0$ are homogeneous steady solution of density and polarization respectively, $\e_x$ is the unit vector along $x$ direction, and $\delta \vw$ and $\delta \rho$ are spatio-temporal perturbations. We linearize \req{eq:Gaussian-W} by writing the equation up to first orders of $\delta \vw$, and $\delta \rho$ (see Appendix~\ref{ap:perturbation}). With the use of Fourier and Laplace transformations we could study the stability of the non-trivial solutions,
\begin{equation}
\delta \rho(\vec{r},t) = \delta \tilde{\rho}(\vec{q},s) e^{st + \im \vec{q}.\vec{r}}, \delta \vw(\vec{r},t) = \delta \tilde{\vw}(\vec{q},s) e^{st + \im \vec{q}.\vec{r}}.
\end{equation}
Here perturbations with eigenvalues $s(q)$ could have non-trivial answers. If $Re[s(q)] > 0$, then the perturbations grow in time and the system is unstable.

The calculation of eigenvalues in non-polar homogeneous state is presented in the Appendix~\ref{ap:non-polar-stability}. Where we show that the homogeneous non-polar state is unstable when $D_r < D_c = \frac{\gamma \rho_0}{2}$. The same prediction is obtained for the truncation method. But as already pointed out previously, the microscopic simulations of the system of  Langevin \reqs{eq:r-dynamics}{eq:theta-dynamics} show strict deviations from the critical values obtained from both continuum theories.

The behavior of polar homogeneous state is more complicated (see Appendix~\ref{ap:polar-stability}), as the polar homogeneous state becomes unstable before the system gets non-polar [see Fig.~\ref{fig:s-map}]. This predicts the formation of traveling bands in the system. Figure~\ref{fig:s-map}, presents the value of $Re[s(\vec{q})]$ in a system with the same parameters as the simulation and integration. One can see that at noise intensity close to $D_r=0.27$, the homogeneous polar solution is unstable which is in agreement with the findings of the numerical integration of the continuum equations in the GA.

\section{Discussion}

In this study we introduced a Gaussian approximation (GA) in order to derive continuum equations for a Vicsek model continuous in time and with ferromagnetic alignment of velocities. It is assumed that the local angle distribution of the particles is a wrapped Gaussian function, and its deviation is related to the local polarization of particles. We used this ansatz to find a closure for Fourier transformation of Fokker-Planck equation that leads to an infinite set of equations. The resulting continuum equations have some differences with the Toner-Tu equations and the usual truncation method described in Refs.~\cite{Bertin2006,Bertin2009}. The nonlinear terms in the GA of continuous-time Vicsek model are of fourth and fifth order of momentum flux, while in the equations of the truncation method, Toner-Tu, and the equations derived by Bertin et al., the non-linearity is of order two and three~\cite{Bertin2006,Bertin2009,Toner1995}. The different non-linearity causes difference in scaling exponent of polar homogeneous solutions near the transition, that for the truncation method we get $\frac{1}{2}$, and for GA we get $\frac{1}{4}$. The same exponent is observed in order parameter of stochastic Kuramoto phase oscillators, when one uses GA~\cite{sonnenschein2013approximate,sonnenschein2013excitable,sonnenschein2014cooperative,Sonnenschein2015}.

We used particle based simulations of the model and extracted orientation distribution of the particles. We observed that the estimated wrapped Gaussian is very close to the simulation observed distribution, providing that  wrapped Gaussian distribution is a reasonable approximation to the distribution of the particles. We also compared the moment generating function of the distributions. The result shows that the distribution is closer to the wrapped Gaussian in high ordered states, and in the low ordered states, it is closer to the distribution predicted by truncation at $k=3$. This tells us that the GA must be more accurate in low noise intensities, while truncation is more closer to the simulation results in high noise intensities.

Moreover, we compared the numeric solutions of the continuum theories. We observed that both continuum models show in agreement with the microscopic simulations a hysteresis behavior. It means that the global polarization differs if we heat or cool the system around the critical noise intensity, where the homogeneous theory predicts the transition between disorder and polar order. Due to the existence of stable inhomogeneous configurations, cooling and heating exhibit different behavior of the global polarization. The found spatio-temporal structures in the GA at different noise levels show the same qualitative behavior as the simulations. The GA global polarization gives closer values in comparison to the truncation method. Further, the GA reaches the simulation values at low noise intensities. Nevertheless, it fails in predicting the correct transition noise intensity, while the truncation continuum theory has a closer, but still different, prediction.

Applying linear stability analysis to the continuum equations in the GA, we found analytically the critical noise intensity for the transition from non-polar to polar order. We also found the noise intensity at which the homogeneous polar order is unstable. These values are in agreement with the results of the numerical integration of the equations of the GA, but differs by the results of the microscopic simulations.

Due to its simplicity and proven accuracy we conclude that the Gaussian approximation is a simple and valid technique for deriving approximative continuum models. It has the ability to reflect qualitatively the behavior of the system in a wide region of noise intensities. In this study, we applied the Gaussian approximation to one of the simplest model of self-propelled particles. Applying it to other models of self-propelled particles~\cite{Menzel2013unidirectional,Romanczuk2016emergent,Menzel2016on,allaei2016,Chen2013universality,Adhyapak2013live}, and whether it works, requires further investigations. However, because of the simplicity and the performance of this method we expect a successful application and functionality to other microscopic models in future investigation.

\begin{acknowledgments}
We are grateful to Bernard Sonnenschein for a stimulating conversation about applying Gaussian theory to active particles, and also for reading the manuscript and his valuable comments. We also thank Fernando Peruani for suggesting the exact mean-field solution, and Ramin Golestanian for valuable discussion. This work was supported by Iran national science foundation (93031724), and 
the Humboldt University (IRTG 1740). Lutz Schimansky-Geier acknowledges support from Humboldt-University at Berlin within the framework of German excellence initiative (DFG).
\end{acknowledgments}

\appendix

\section{Simplification of Continuum equations}
\label{ap:simplification}

Like for the truncation method (Sec.\ref{subsection:truncation}), there exist the scaling behaviors $\vw \sim \epsilon$, $\partial_t \sim \epsilon$, and $\grad \sim \epsilon$ close to the transition~\cite{Bertin2009}. We rewrite \req{eq:Gaussian-W} up to $\epsilon^5$ to obtain a simplified equation, that is
\begin{equation}
\begin{aligned}
&\Dif{{\vw}}{t} + \grad. \tensor{\mathcal{J}}_{\vw} = \left[ \frac{\gamma \rho}{2} - D_r - \frac{\gamma W^4}{2 \rho^3} \right] \vw \\
&- \frac{v_0}{2} \grad \left[  \rho - \frac{W^4}{\rho^3} \right] + \left[ \frac{\gamma R^2}{16} \rho + K \right] \nabla^2 \vw.
\end{aligned}
\label{eq:Gaussian-W2}
\end{equation}

Because the GA function is better deep in ordered state, we did not integrate \req{eq:Gaussian-W2} numerically. 

\section{Perturbation}
\label{ap:perturbation}

We consider a system at homogeneous steady state solution. Without loss of generality, the polarization can be along $x$ direction, i.e. $\vw_0 = \w_0 \e_x$. The linearized version of \reqs{eq:continuity}{eq:Gaussian-W} around the homogeneous steady state solution  is as the following
\begin{equation}
\Dif{\delta \rho}{t} + v_0 \grad \cdot \delta \vw - K \nabla^2 \delta \rho = 0,
\label{eq:Gaussian-drho}
\end{equation}
and
\begin{equation}
\begin{aligned}
&\Dif{\delta {\vw}}{t} + v_0 \p_0^3 \left[ 2 \partial_x \delta \w_x \e_x +  \grad \cdot \delta \vw \e_x + \partial_x \delta \vw \right] \\
&- 3 v_0 \p_0^4 \partial_x \delta \rho \e_x = \left[ \frac{\gamma}{2} \left( 1 + 3 \p_0^4 \right) \delta \rho - 2 \gamma \p_0^3 \delta \w_x \right] \w_0 \e_x \\
&+\left[ \frac{\gamma \rho_0}{2} \left( 1 - \p_0^4 \right) - D_r  \right] \delta \vw \\
&- \frac{v_0}{2} \grad \left[ \left( 1 + 3 \p_0^4 \right) \delta \rho - 4 \p_0^3 \delta \w_x \right] \\
&+ \frac{\gamma R^2 \rho_0}{16} \left[ \left( 1+\p_0^4 \right) \nabla^2 \delta \vw - 2 \p_0^4 \e_x \nabla^2 \delta \w_x \right] \\
&+ K \nabla^2 \delta \vw
\end{aligned}
\label{eq:Gaussian-dw}
\end{equation}

A Fourier transformation in space and a Laplace transformation in time gives the following,
\begin{equation}
s \delta \tilde{\rho} + \im v_0 \vec{q}  \cdot \delta \tvw + K q^2 \delta \tilde{\rho} = 0,
\label{eq:Gaussian-drho-fourier}
\end{equation}
\begin{equation}
\begin{aligned}
&s \delta \tvw + \im v_0 \p_0^3 \left[ 2 q_x \delta \tw_x \e_x +  \vec{q} \cdot \delta \tvw \e_x + q_x \delta \tvw \right] \\
&- 3 \im v_0 \p_0^4 q_x \delta \tilde{\rho} \e_x = \left[ \frac{\gamma}{2} \left( 1 + 3 \p_0^4 \right) \delta \tilde{\rho} - 2 \gamma \p_0^3 \delta \tw_x \right] \w_0 \e_x \\
&+\left[ \frac{\gamma \rho_0}{2} \left( 1 - \p_0^4 \right) - D_r  \right] \delta \tvw \\
&- \frac{\im v_0}{2} \vec{q} \left[ \left( 1 + 3 \p_0^4 \right) \delta \tilde{\rho} - 4 \p_0^3 \delta \tw_x \right] \\
&- \frac{\gamma R^2 q^2 \rho_0}{16} \left[ \left( 1+\p_0^4 \right) \delta \tvw - 2 \p_0^4 \e_x \delta \tw_x \right] \\
&- K q^2 \delta \tvw.
\end{aligned}
\label{eq:Gaussian-dw-fourier}
\end{equation}

\section{Stability of non-polar state}
\label{ap:non-polar-stability}

Using the non polar homogeneous answer $\w_0 = 0$ in \reqs{eq:Gaussian-dw-fourier}{eq:Gaussian-drho-fourier} one finds the simple linearized equations,
\begin{equation}
\begin{aligned}
&\left[ s - \frac{\gamma \rho_0}{2} + \frac{\gamma \rho_0 R^2  q^2}{16} + D_r + K q^2 \right]  \delta \tilde{\vw} \\
&+ \im \frac{v_0}{2} \vec{q} \delta \tilde{\rho} = 0,
\end{aligned}
\end{equation}
and
\begin{equation}
\left( s + K q^2 \right) \delta \tilde{\rho} + \im v_0 \vec{q} . \delta \tilde{\vw} = 0.
\end{equation}

We substitute the last two equations into one another to find,
\begin{equation}
\begin{aligned}
&\Bigg[ \left( s + K q^2 \right) \left( s - \frac{\gamma \rho_0}{2} + \frac{\gamma \rho_0 R^2 q^2}{16} + D_r + K q^2 \right) + q^2 \frac{v_0^2}{2}  \Bigg] \\
&\times \delta \tilde{\rho} = 0.
\end{aligned}
\end{equation}
For non trivial answers ($\delta \rho \neq 0$), the coefficient of $\delta \tilde{\rho}$ must be equal to zero. This gives us value of $s$ as
\begin{equation}
\begin{aligned}
s_{\pm} &= - K q^2 + \frac{1}{2} \Bigg[ -\left( D_r - \frac{\gamma \rho_0}{2} + \frac{\gamma \rho_0 R^2 q^2}{16} \right) \\
&\pm \sqrt{\left( D_r - \frac{\gamma \rho_0}{2} + \frac{\gamma \rho_0 R^2 q^2}{16} \right)^2 - 2 v^2_0 q^2 } \Bigg].
\end{aligned}
\end{equation}

The non-polar state is stable as long as $Re[s] < 0$, and this is satisfied as long as $D_r > D_c = \frac{\gamma \rho_0}{2}$.

\section{Stability of longitudal fluctuation of polar state}
\label{ap:polar-stability}

For the homogeneous polar state with $\p_0 = [1 - 2 D_r / (\gamma \rho_0)]^\frac{1}{4}$ we consider the longitude perturbations. Therefore, $\delta \vw = \delta \w \e_x$, and $\vec{q}=q \e_x$. Using the identity $\gamma \rho_0 (1-\p_0^4)/2 = D_r$, and $\w_0 = \rho_0 \p_0$, the longitudinal perturbations in \reqs{eq:Gaussian-drho-fourier}{eq:Gaussian-dw-fourier} result to the following equations
\begin{equation}
s \delta \tilde{\rho} + \im v_0 q  \delta \tw + K q^2 \delta \tilde{\rho} = 0,
\end{equation}
and
\begin{equation}
\begin{aligned}
&\Big[ s + K q^2 + 2 \gamma \rho_0  \p_0^4 + 2 \im v_0 q \p_0^3 + \frac{D_r R^2 q^2}{8}  \Big] \delta \tw \\
&+ \left[ \frac{\im v_0 q}{2} \left( 1 - 3 \p_0^4 \right) - \frac{\gamma \rho_0}{2} \left( 1 + 3 \p_0^4 \right) \p_0  \right] \delta \tilde{\rho}  = 0.
\end{aligned}
\end{equation}

To have nontrivial solutions, the $s$ must satisfy the following equation
\begin{equation}
\begin{aligned}
&\Big[ s + K q^2 + 2 \gamma \rho_0 \p_0^4 + 2 \im v_0 q \p_0^3 \\
&+ \frac{D_r R^2 q^2}{8}  \Big] \left[ s + K q^2 \right] \\
&- \im v_0 q \left[ \frac{\im v_0 q}{2} \left( 1 - 3\p_0^4 \right) - \frac{\gamma \rho_0}{2} \left( 1 + 3 \p_0^4 \right) \p_0  \right]  = 0,
\end{aligned}
\end{equation}
with the solutions
\begin{equation}
\begin{aligned}
&s = - Kq^2 -\gamma \rho_0 \p_0^4 - \im v_0 q \p_0^3 - \frac{D_r R^2 q^2}{16} \\
&\pm \Bigg[ \Big( \gamma \rho_0 \p_0^4 + \im v_0 q \p_0^3 + \frac{D_r R^2 q^2}{16} \Big)^2 \\
&+ \frac{\im v_0 q}{2} \Big( \im v_0 q (1 - 3\p_0^4) - \gamma \rho_0 (1 + 3 \p_0^4) \p_0 \Big)  \Bigg]^\frac{1}{2}
\end{aligned}
\label{eq:s-polar}
\end{equation}
where $x = 2 \gamma \rho_0 \p_0^4$, and $\p_0 = (1 - \frac{2 D_r}{\gamma \rho_0})^{\tfrac{1}{4}}$. Equation~(\ref{eq:s-polar}) has a positive result for small $q$ when noise is higher than a threshold [see Fig.~\ref{fig:s-map}]. This shows the emergence of an inhomogeneous answer which is the polar band structure.

\bibliography{references}

\end{document}